\documentclass[11pt,a4paper]{article}
\def\bea{\begin{eqnarray}}
\def\eea{\end{eqnarray}}
\def\nn{\nonumber}
\def\beq{\begin{equation}}
\def\eeq{\end{equation}}
\def\ba{\beq\new\begin{array}{c}}
\def\ea{\end{array}\eeq}
\def\be{\ba}
\def\ee{\ea}

\makeatletter
\newdimen\normalarrayskip              % skip between lines
\newdimen\minarrayskip                 % minimal skip between lines
\normalarrayskip\baselineskip
\minarrayskip\jot
\newif\ifold             \oldtrue            \def\new{\oldfalse}
\def\arraymode{\ifold\relax\else\displaystyle\fi} % mode of array entries
\def\eqnumphantom{\phantom{(\theequation)}}     % right phantom in eqnarray
\def\@arrayskip{\ifold\baselineskip\z@\lineskip\z@
     \else
     \baselineskip\minarrayskip\lineskip2\minarrayskip\fi}
\def\@arrayclassz{\ifcase \@lastchclass \@acolampacol \or
\@ampacol \or \or \or \@addamp \or
   \@acolampacol \or \@firstampfalse \@acol \fi
\edef\@preamble{\@preamble
  \ifcase \@chnum
     \hfil$\relax\arraymode\@sharp$\hfil
     \or $\relax\arraymode\@sharp$\hfil
     \or \hfil$\relax\arraymode\@sharp$\fi}}
\def\@array[#1]#2{\setbox\@arstrutbox=\hbox{\vrule
     height\arraystretch \ht\strutbox
     depth\arraystretch \dp\strutbox
     width\z@}\@mkpream{#2}\edef\@preamble{\halign
\noexpand\@halignto
\bgroup \tabskip\z@ \@arstrut \@preamble \tabskip\z@ \cr}%
\let\@startpbox\@@startpbox \let\@endpbox\@@endpbox
  \if #1t\vtop \else \if#1b\vbox \else \vcenter \fi\fi
  \bgroup \let\par\relax
  \let\@sharp##\let\protect\relax
  \@arrayskip\@preamble}
\def\eqnarray{\stepcounter{equation}%
              \let\@currentlabel=\theequation
              \global\@eqnswtrue
              \global\@eqcnt\z@
              \tabskip\@centering
              \let\\=\@eqncr
              $$%
 \halign to \displaywidth\bgroup
    \eqnumphantom\@eqnsel\hskip\@centering
    $\displaystyle \tabskip\z@ {##}$%
    \global\@eqcnt\@ne \hskip 2\arraycolsep
         $\displaystyle\arraymode{##}$\hfil
    \global\@eqcnt\tw@ \hskip 2\arraycolsep
         $\displaystyle\tabskip\z@{##}$\hfil
         \tabskip\@centering
    &{##}\tabskip\z@\cr}
\begingroup\ifx\undefined\newsymbol \else\def\input#1 {\endgroup}\fi
\input amssym.def \relax
\input amssym
\newfont{\hr}{msbm10}
\newfont{\ams}{msam10}
\font\numbers=cmss12
\font\upright=cmu10 scaled\magstep1
\def\stroke{\vrule height8pt width0.4pt depth-0.1pt}
\def\topfleck{\vrule height8pt width0.5pt depth-5.9pt}
\def\botfleck{\vrule height2pt width0.5pt depth0.1pt}
\def\Zmath{\vcenter{\hbox{\numbers\rlap{\rlap{Z}\kern 0.8pt\topfleck}\kern 2.2pt
                   \rlap Z\kern 6pt\botfleck\kern 1pt}}}
\def\Qmath{\vcenter{\hbox{\upright\rlap{\rlap{Q}\kern
                   3.8pt\stroke}\phantom{Q}}}}
\def\Nmath{\vcenter{\hbox{\upright\rlap{I}\kern 1.7pt N}}}
\def\Cmath{\vcenter{\hbox{\upright\rlap{\rlap{C}\kern
                   3.8pt\stroke}\phantom{C}}}}
\def\Rmath{\vcenter{\hbox{\upright\rlap{I}\kern 1.7pt R}}}
\def\Z{\ifmmode\Zmath\else$\Zmath$\fi}
\def\Q{\ifmmode\Qmath\else$\Qmath$\fi}
\def\N{\ifmmode\Nmath\else$\Nmath$\fi}
\def\C{\ifmmode\Cmath\else$\Cmath$\fi}
\def\R{\ifmmode\Rmath\else$\Rmath$\fi}
\newcounter{app}

\def\app{\setcounter{equation}{0}
\def\theequation{\Alph{app}.\arabic{equation}}\par
   \addvspace{4ex}
   \@afterindentfalse
  \secdef\@app\@dapp}

\newcommand\@app{\@startsection {app}{1}{0ex}%
                                   {-3.5ex \@plus -1ex \@minus -.2ex}%
                                   {2.3ex \@plus.2ex}%
                                   {\normalfont\Large\bf}}
\def\@dapp#1{%
{\parindent \z@ \raggedright  \bf #1}\par\nobreak}
\def\l@app#1#2{\ifnum \c@tocdepth >\z@
    \addpenalty\@secpenalty
    \addvspace{1.0em \@plus\p@}%
    \setlength\@tempdima{8em}%
    \begingroup
      \parindent \z@ \rightskip \@pnumwidth
      \parfillskip -\@pnumwidth
      \leavevmode \bfseries
      \advance\leftskip\@tempdima
      \hskip -\leftskip
      #1\nobreak\hfil \nobreak\hb@xt@\@pnumwidth{\hss #2}\par
    \endgroup\fi}
\newcounter{sapp}[app]

\def\sapp{\def\theequation{\Alph{app}.\arabic{equation}}
\par
\@afterindentfalse
  \secdef\@sapp\@dsapp}
\newcommand{\@sapp}{\@startsection{sapp}{2}{\z@}%
                                     {-3.25ex\@plus -1ex \@minus 
-.2ex}%
                                     {1.5ex \@plus .2ex}%
                                     {\normalfont\large\bfseries}}

\def\@dsapp#1{%
{\parindent \z@ \raggedright  \bf #1
}\par\nobreak}
\newcommand{\l@sapp}{\@dottedtocline{2}{1.5em}{2.3em}}

\catcode`\@=11

\def\section{\@startsection{section}{1}{\z@}{3.5ex plus 1ex minus
   .2ex}{2.3ex plus .2ex}{\large\bf}}

\def\thesection{\Roman{section}.}

\def\appendix{\setcounter{section}{0}
        \def\thesection{Appendix }
       \def\theequation{\Alph{section}.\arabic{equation}}}

\@addtoreset{equation}{section}
\def\theequation{\arabic{section}.\arabic{equation}}
\def\2{{1\over 2}}
\def\N2{${\cal N}=2$}

\def\be{ \begin{eqnarray} }
\def\ee{ \end{eqnarray} }

\def\bea{\begin{eqnarray}}
\def\eea{\end{eqnarray}}
\def\nn{\nonumber}

\def\beq{\begin{equation}}
\def\eeq{\end{equation}}
\def\ba{\beq\new\begin{array}{c}}
\def\ea{\end{array}\eeq}
\def\be{\ba}
\def\ee{\ea}

\title{Gluino-Condensate (CIV-DV) Prepotential  \\
          from its Whitham-Time Derivatives}

\author{
H.Itoyama$^{1}$ and A.Morozov$^{2}$ \
\\ \normalsize \em $^{1}$ 
Department of Mathematics and Physics, Osaka City University, Japan
\\
\normalsize \em $^{2}$
Institute of Theoretical and Experimental Physics, Moscow}

\date{January, 2003}

\begin{document}

\maketitle

\vspace{-7.7cm}

\begin{center}
\hfill hep-th/0301136 \\
\hfill OCU-PHYS 197\\
\hfill ITEP/TH-07/03
\end{center}

\vspace{6.5cm}

\begin{abstract}

We describe an expedient way to derive the CIV-DV prepotential
 in power series expansion in $S_i$. This is
based on integrations of equations for its derivatives
$\partial {\cal F}/\partial T_m$ with respect to additional
(Whitham) moduli $T_m$.
For illustrative purposes, we calculate explicitly the
leading terms of the expansion and explicitly check
some components of
the WDVV equations to the leading order.
Extension to any higher order is simple and algorithmic.

\end{abstract}

\vspace{2.0cm}

%\end{titlepage}

\bigskip

\section{Introduction}

This paper continues discussion of the CIV-DV prepotential
\cite{CIV,DV,followup} from the perspective of the
Seiberg-Witten theory \cite{SW,intfollowup}, originated in \cite{IM4}.
We explained in \cite{IM4} that the full set of moduli in DV model
is $2n$-dimensional (rather than $n$-dimensional),
and the {\it flat} coordinates on it are

\be
S_k = \oint_{A_k} dS_{DV},\ \ \
T_k = \ res_\infty\ x^{-k} dS_{DV},\ \ \
k=1,\ldots,n  \;\;,
\label{moduli}
\ee
where

\be
dS_{DV}(x) = Y(x)dx, \ \ \ \nn \\
Y^2(x)  = g^2P_n^2(x) + f_{n-1}(x)
 = g^2P_n^2(x) + 2gP_n(x)\sum_{i=1}^n\frac{\tilde S_i}{x-\alpha_i}, \nn \\
P_n(x) = \prod_{i=1}^n (x-\alpha_i) = \sum_{m=0}^n u_m x^m,\ \ \
u_n =1, \ \ u_m = (-)^{n-m} e_{n-m}(\alpha),
\label{dS}
\ee
and

\be
e_m(\alpha) = \sum_{i_1<\ldots <i_m}\alpha_{i_1}\ldots \alpha_{i_m}
\ee
are symmetric polynomials in $\alpha$'s.
As $dS_{DV} = (gP(x) + O(x^{-1}))dx$ at $x \rightarrow \infty$,
the Whitham times are obviously

\be
T_m = gu_{m-1},\ \ m=1,\ldots,n.
\label{T=u}
\ee

In \cite{IM4,IM5} an additional restriction
$T_n = gu_{n-1} = -g \displaystyle{\sum_{i=1}^n\alpha_i = 0}$
was imposed on the moduli space, (so that only {\it holomorphic}
differentials arise in discussion of the {\it regularized} DV model).
However, as correctly pointed out in \cite{Chetal}, this constraint
can break some nice properties of the theory: in particular, it leads
to complications with the WDVV equations \cite{WDVV,MMM}, which are observed
in \cite{IM5}. If $T_n$ is included back into the set of moduli, the WDVV
equations were shown in \cite{Chetal} to hold for the CIV-DV
prepotential in the most straightforward way (with the residue
formula, suggested in \cite{IM4}).\footnote{
In order to obtain this result, in eq.(35) of
ref.\cite{IM4} one should not transfer from
$\sum_{dS_{DV}=0}  \cdots $ to $\sum_{d{\cal P}_{2n}=0} \cdots$
(in this transition one should be careful about contributions
from the zeroes of ${\cal R}_{2n}$)
and, accordingly, in (45) and (46) of \cite{IM4}
an algebra of polynomials modulo ${\cal P}_{2n}$
(rather than ${\cal P}'_{2n}$) should be considered.
For {\it such} algebra to be closed, one needs polynomials
of degree $2n-1$ rather than $2n-2$: the matching
condition is always that
 "(degree of polynomials that form the algebra) $=$
 (degree of polynomial which defines factorization) $-1$.
Still, the {\it experimental} result of \cite{IM5},
concerning the validity of WDVV equations for a whole family of the CIV-DV-like
prepotentials
with just {\it three} moduli $S_1,S_2,T$ for $n=2$, remains true.
Moreover, by now we checked this fact
with the help of Maple and it appears that, at least up to the order $S^5$
the WDVV equations are satisfied whenever $\nu = \pm\frac{1}{b\mp 1}$.
Moreover, this
result does not depend on the coefficients in front of the $S^3$, $S^4$ and
$S^5$
terms. The result remains theoretically unexplained at this moment.
}

Now, after the origin of difficulties in ref.\cite{IM4} is
identified and eliminated (and thus the whole approach is justified),
we can continue developing the theory along the lines of that paper.
The task of the present text is modest:
we develop a calculus, based on expansion of $dS_{DV}$ in
 power series  in $\tilde S_i$:

\be
dS_{DV}(x) = gP_n(x)dx + \sum_{i=1}^n \tilde S_i\frac{dx}{x-\alpha_i} -
%\nn \\ -
\frac{1}{2g}\sum_{i,j=1}^n \tilde S_i\tilde
S_j\frac{dx}{(x-\alpha_i)(x-\alpha_j)P_n(x)} - \ldots
\label{dSexpansion}
\ee
Its main advantage is
that  contour integrals are substituted by residues
at points $x=\alpha_i$ and $x=\infty$. The equations

\be
\frac{\partial{\cal F}_{DV}}{\partial S_k} = \int_{B_k} dS_{DV}
\label{derFS}
\ee
are difficult to handle.
The problem is that particular terms of expansion (\ref{dSexpansion})
for $dS_{DV}$ are singular at $x = \alpha_i$
while $dS_{DV}$ itself has no singularity at this point.
The integral $B_k$ in (\ref{derFS}) actually goes between some
large $\Lambda$ and $\tilde\alpha_i$, which is a root of
$Y^2(x)$, very close to $\alpha_i$ at small $\tilde S_i$:
$\tilde\alpha_i-\alpha_i \sim \sqrt{\tilde S_i}$.
Thus after the integration of particular terms in (\ref{dSexpansion}),
$\tilde S$ can emerge in the denominator, and this breaks the naive
structure of the power expansion in $\tilde S_i$ and makes the evaluation of
$\partial{\cal F}_{DV}/\partial S_i$ more sophisticated. See \cite{IM6}.

However, one can instead use a set of the remaining equations\footnote{Additional
$\Lambda$-dependent term in this definition is important to
reproduce the prepotential, associated with the spectral curve.
If this term is omitted from (\ref{derFT}), one obtains the
planar-matrix-model prepotential, different from (\ref{prepo})
by the absence of terms with $\displaystyle{\sum_{i=1}^n S_i W_{n+1}(\Lambda)}$ 
and $\displaystyle{(\sum_{i=1}^n S_i)^2 \log\Lambda}$.
This difference is rarely important, but sometimes it is:
for example the Ward identities in section \ref{WId} below
are derived in the presence of the $\Lambda$-dependent terms.
}
\be
\frac{\partial{\cal F}_{DV}}{\partial T_m} =
\frac{1}{m} res_\infty\ (x^m - \Lambda^m) dS_{DV}
%-???\Lambda^{k}\sum_{j=1}^n S_j
=\frac{1}{m}\sum_{i=1}^n (\alpha_i^m - \Lambda^m)\tilde S_i
%-???\Lambda^{k}\sum_{j=1}^n S_j
\label{derFT}
\ee
to define the prepotential, which can be easily integrated
for any given power of $S_i$ and this procedure provides
the CIV-DV prepotential as a power series in $S_i$
with coefficients made from $\alpha_i$,\footnote{Eq.(\ref{prepo})
differs from analogous expressions in refs.\cite{IM4,IM6} by rescalings
$$
S_i = \frac{1}{2\pi i}S_{i \cite{IM6}}, \ \
\frac{\partial{\cal F}}{\partial S_i} = \frac{1}{2}
\frac{\partial{\cal F}_{\cite{IM6}}}{\partial S_{i \cite{IM6}}},
\ \ i.e. \ {\cal F} = \frac{1}{4\pi i} {\cal F}_{\cite{IM6}}
$$
This is because, in the present paper, we include the factor
$(2\pi i)^{-1}$ into the definition of contour integrals in
(\ref{moduli}) and omit factor  $2$ in eqs.(\ref{derFS})
and (\ref{derFT}). (Such factor would appear in (\ref{derFT})
to account for the fact that $x=\infty$ describes two points
on the two sheets of the hyperelliptic spectral curve. As it is
omitted  in (\ref{derFT}), there should be no $2$ in (\ref{derFS}) as well).
}

\be
{\cal F}_{DV}(S|\alpha) =
g\sum_{i=1}^n S_i ( W_{n+1}(\alpha_i) - W_{n+1}(\Lambda)) + \nn \\
- \frac{1}{2}(\sum_{i=1}^n S_i)^2 \log\Lambda
+\frac{1}{4}\sum_{i+1}^n S_i^2( \log S_i - \frac{3}{2})
-\frac{1}{4}\sum_{i<j} (S_i^2 - 4S_iS_j + S_j^2)\log\alpha_{ij} + \nn \\
+ \sum_{p=3}^\infty g^{2-p} {\cal F}_p(S|\alpha),
\label{prepo}
\ee
with $W_{n+1}'(x) = P_n(x)$, i.e.

\be
W_{n+1}(x) = \sum_{k=0}^n\frac{u_kx^{k+1}}{k+1} \;\;.
\label{W}
\ee
Actually, the two $\alpha$-independent terms in the second line of
eq.(\ref{prepo}) cannot be found from (\ref{derFT}), but they can
be easily fixed by other methods.
Note also that in all orders of the $\tilde S$-expansion

\be
\sum_{i=1}^n \tilde S_i = \sum_{i=1}^n  S_i
\ee
since the sum of the integrals over all of the  $A_i$ contours is equal to
residue at infinities, $x=\infty$.

This method provides considerable
simplification  when evaluating higher corrections to the prepotential,
making calculations comparably simple to those in matrix models
(while straightforward evaluation involving contour integrals
is incredibly sophisticated. See \cite{IM6}).
This allows us to check general
arguments and proofs of \cite{prepth,IM4,Chetal} (but necessarily
involving transcendental expressions)
concerning the properties of the prepotential and the WDVV equations,
with explicit and elementary calculations -- for particular terms in
the series expansion.

What is more important, this method provides a non-transcendental
definition of particular terms ${\cal F}_p(S|\alpha)$ in the prepotential
expansion, which are rational functions of $\alpha$'s and thus
are expressed through rational (rather than hyperelliptic) integrals.
Such representation is useful for comparison to matrix-model
calculations and instanton calculus \cite{NeF}.

The very simple
form of the prepotential $T$-derivatives (\ref{derFT})
deserves to be mentioned:
the r.h.s. is just a linear function of $\tilde S_i$ (though,
of course, it becomes an infinite series when expressed in terms
of the {\it flat} moduli $S_i$)

 In the next section, we derive
a relation between $\tilde S_i$ and $S_i$
and sum rules for rational functions of $\alpha_i$ associated with this. 
The $T$-derivatives of the roots $\alpha_i$ are computed
in section three, and
  consistency  is checked on
the second derivatives of the prepotential in section four.
In section five, we describe how to obtain
 the prepotential from its $T$-derivatives, and
in section six, ${\cal L}_{-1}$ and ${\cal L}_{0}$-constraints
 on ${\cal F}_{DV}$ are given.
In section seven and eight, we discuss
 the WDVV equations of the prepotential, and check them
 explicitly to the leading order.

\section{Relation between $\tilde S_i$ and $S_i$
and sum rules for rational functions of $\alpha_i$}

%Parameterization (\ref{dS}) is very convenient for evaluation of
%residues at infinities $x=\infty$.
Parameters $\tilde S_i$ can
be expressed through $S_i$ by taking residues at $x=\alpha_i$.
In more detail, from (\ref{dSexpansion}):

\be
dS_{DV} = gP_n(x)dx + \sum_{i=1}^n \tilde S_i\frac{dx}{x-\alpha_i} -
%\nn \\ -
\frac{1}{2g}\sum_{j,k=1}^n \tilde S_j\tilde S_k
\frac{dx}{(x-\alpha_j)(x-\alpha_k)P_n(x)} - \ldots
\label{dSexpansion1}
\ee
Then

\be
S_i = \ res_{\alpha_i}\ dS_{DV} =
\tilde S_i + \frac{1}{2g}\eta_{i;jk}\tilde S_j\tilde S_k + \ldots
\label{SvstildeS}
\ee
or
\be
\tilde S_i =
S_i - \frac{1}{2g}\eta_{i;jk}S_jS_k + \ldots
\label{tildeSvsS}
\ee

In order to evaluate the coefficients $\eta_{i;jk}$ one needs the following
integrals. (As usual, the factor of $(2\pi i)^{-1}$ is included into the
definition of contour integral.)

\be
\oint_{\alpha_i} \frac{dx}{(x-\alpha_j)(x-\alpha_k)P_n(x)} =
\frac{1}{\alpha_{ij}\alpha_{ik}\Delta_i},\ \ {\rm for} \ \ j,k\neq i;
\nn \\
\oint_{\alpha_i} \frac{dx}{(x-\alpha_i)(x-\alpha_j)P_n(x)} =
\oint_0 \frac{dz}{z^2(\alpha_{ij}+z)^2\prod_{k\neq i,j}(\alpha_{ik}+z)}
= \nn \\
= -\frac{1}{\alpha_{ij}\Delta_i}\left(\frac{2}{\alpha_{ij}} +
\sum_{k\neq i,j}\frac{1}{\alpha_{ik}}\right), \ \ {\rm for} \ \ i\neq j; \nn \\
\oint_{\alpha_i} \frac{dx}{(x-\alpha_i)^2P_n(x)} =
\oint_0 \frac{dz}{z^3\prod_{j\neq i}(\alpha_{ij}+z)}
%= \nn \\
= \frac{1}{\Delta_i}\left(\sum_{j\neq i}\frac{1}{\alpha_{ij}^2} +
\sum_{\stackrel{j<k}{j,k\neq i}}\frac{1}{\alpha_{ij}\alpha_{ik}}\right)
\label{ress}
\ee
Here $\alpha_{ij} = \alpha_i - \alpha_j$,
$\Delta_i = \prod_{j\neq i}\alpha_{ij}$.
Using the fact that the sums over all residues
 of the integrand should vanish, we obtain from
(\ref{ress}) the following identities (sum rules):\footnote{This proof of
(\ref{ide}) was suggested by V.Pestun.
The first of these identities played an important role in ref.\cite{IM6}
(eq.(4.11) of that paper). The second identity can be used to convert
the $S_i^3$ contributions to the prepotential, found in \cite{IM6}  as they pull
$\frac{1}{\Delta_j}$ from inside the summation. This plays an important role for
comparing, say, with matrix model calculations.}

\be
\sum_{k\neq i,j}\frac{1}{\alpha_{ik}\alpha_{jk}\Delta_k} =
\frac{2}{\alpha_{ij}^2}\left(\frac{1}{\Delta_i} + \frac{1}{\Delta_j}\right) +
\frac{1}{\alpha_{ij}}\sum_{k\neq i,j} \left(\frac{1}{\alpha_{ik}\Delta_i} -
\frac{1}{\alpha_{jk}\Delta_j}\right), \nn \\
\sum_{j\neq i}\frac{1}{\alpha_{ij}^2\Delta_j} = -
\frac{1}{\Delta_i}\sum_{j\neq i}\frac{1}{\alpha_{ij}^2} -
\frac{1}{\Delta_i}\sum_{\stackrel{j<k}{j,k\neq i}}\frac{1}{\alpha_{ij}\alpha_{ik}}
\label{ide}
\ee

In what follows we will need
also a slight generalization of these sum rules, with an extra factor of $x^m$
in the integrand (with $m\leq n$ to avoid contributions from $x = \infty$):

\be
\oint_{\alpha_i} \frac{x^m dx}{(x-\alpha_i)(x-\alpha_j)P_n(x)} =
\oint_0 \frac{(\alpha_i +z)^m dz}
{z^2(\alpha_{ij}+z)^2\prod_{k\neq i,j}(\alpha_{ik}+z)}
= \nn \\
= -\frac{\alpha_i^m}{\alpha_{ij}\Delta_i}\left(\frac{2}{\alpha_{ij}} +
\sum_{k\neq i,j}\frac{1}{\alpha_{ik}}\right)
+ \frac{m\alpha_i^{m-1}}{\alpha_{ij}\Delta_i}, \ \ {\rm for}\ i\neq j;
\nn \\
\oint_{\alpha_i} \frac{x^m dx}{(x-\alpha_i)^2 P_n(x)} =
\oint_0 \frac{(\alpha_i +z)^m dz}
{z^3\prod_{j\neq i}(\alpha_{ij}+z)}
= \nn \\
= \frac{\alpha_i^m}{\Delta_i}
\left(\sum_{j\neq i}\frac{1}{\alpha_{ij}^2} +
\sum_{\stackrel{j<k}{j,k\neq i}}\frac{1}{\alpha_{ij}\alpha_{ik}}\right)
- \frac{m\alpha_i^{m-1}}{\Delta_i}\sum_{j\neq i}\frac{1}{\alpha_{ij}}
+ \frac{m(m-1)}{2}\frac{\alpha_i^{m-2}}{\Delta_i}
\ee
and, as corollaries,

\be
\sum_{k\neq i,j}\frac{\alpha_k^m}{\alpha_{ik}\alpha_{jk}\Delta_k} =
\frac{2}{\alpha_{ij}^2}\left(\frac{\alpha_i^m}{\Delta_i} +
\frac{\alpha_j^m}{\Delta_j}\right) +
\frac{1}{\alpha_{ij}}\sum_{k\neq i,j} \left(\frac{\alpha_i^m}{\alpha_{ik}\Delta_i} -
\frac{\alpha_j^m}{\alpha_{jk}\Delta_j}\right) - \nn\\ -
\frac{m}{\alpha_{ij}}\left(\frac{\alpha_i^{m-1}}{\Delta_i} -
\frac{\alpha_j^{m-1}}{\Delta_j}\right)
\label{idem}
\ee
and

\be
\sum_{j\neq i}\frac{\alpha_j^m}{\alpha_{ij}^2\Delta_j} =
- \frac{\alpha_i^m}{\Delta_i}
\left(\sum_{j\neq i}\frac{1}{\alpha_{ij}^2} +
\sum_{\stackrel{j<k}{j,k\neq i}}\frac{1}{\alpha_{ij}\alpha_{ik}}\right)
+ \nn \\
+ \frac{m\alpha_i^{m-1}}{\Delta_i}\sum_{j\neq i}\frac{1}{\alpha_{ij}}
- \frac{m(m-1)}{2}\frac{\alpha_i^{m-2}}{\Delta_i} \;\;. 
\label{idemm}
\ee
A simpler sum rule, associated with the integrand
$\frac{x^m dx}{(x-\alpha_i)P_n(x)}$,
states:

\be
\sum_{j\neq i}\frac{\alpha_j^m}{\alpha_{ij}\Delta_j} =
-\frac{\alpha_i^m}{\Delta_i}\sum_{j\neq i}\frac{1}{\alpha_{ij}} +
\frac{m\alpha_i^{m-1}}{\Delta_i},\ \ m<n
\label{ideem}
\ee

Coming back to our problem, one can read from (\ref{ress}) expressions
for the coefficients $\eta_{i;jk}$ in (\ref{SvstildeS}):

\be
\eta_{i;ii} = -\frac{1}{\Delta_i}
\left(\sum_{j\neq i}\frac{1}{\alpha_{ij}^2} +
\sum_{\stackrel{j<k}{j,k\neq i}}\frac{1}{\alpha_{ij}\alpha_{ik}}\right)
\ \stackrel{eq.(\ref{ide})}{=}\
\sum_{j\neq i}\frac{1}{\alpha_{ij}^2\Delta_j}, \nn \\
\eta_{i;ij} =
\frac{1}{\Delta_i}\left(\frac{2}{\alpha_{ij}^2} +
\frac{1}{\alpha_{ij}}\sum_{k\neq i,j}\frac{1}{\alpha_{ik}}\right), \nn \\
%\eta_{i;jj} = -\frac{1}{\alpha_{ij}^2\Delta_i}, \nn \\
\eta_{i;jk} = -\frac{1}{\alpha_{ij}\alpha_{ik}\Delta_i},\ \ {\rm for}\ j,k\neq i.
\label{eta}
\ee
These coefficients satisfy the sum rules

\be
\eta_{i;ii} = -\frac{1}{2}\sum_{j\neq i} \eta_{i;ij} =
-\sum_{j\neq i}\eta_{j;ii}, \nn \\
\eta_{i,jj} =
-\frac{1}{2}\left(\eta_{i;ij} + \sum_{k\neq i,j}\eta_{i;jk}\right)
 = -\frac{1}{2}\sum_{k\neq j}\eta_{i;jk}; \nn \\
\sum_{i=1}^n \eta_{i;jk} = 0, \ \ \forall j,k.
\label{sumruleseta}
\ee
To prove the last of these sum rules for $j\neq k$ one should apply (\ref{ide}).
This sum rule guarantees that
$\displaystyle{\sum_{i=1}^n \tilde S_i = \sum_{i=1}^n  S_i}$ to quadratic order
in $\tilde S^2$.

\section{$T$-derivatives of the roots $\alpha_i$}

To obtain ${\cal F}$ as a function of $S_i$ and $T_m$, we need to express
$\alpha$'s in (\ref{prepo}) through $T_m$.
The modulus $T_{m+1} = gu_m = (-)^{n-m} ge_{n-m}(\alpha)$
is a symmetric polynomial (of degree $m$) in $\alpha_i$. Conversely, $\alpha_i$
is a section of an $n$-dimensional bundle over the space of $T_m$'s,
(that is, $\alpha_i$'s are obtained as solution of a system of algebraic equations
and different $\alpha_i$'s are considered as different roots and get interchanged
 when $T_m$'s move around the singularities).

The $T$($u$)-derivatives of $\alpha$'s can be obtained by inversion of the
matrix

\be
\frac{\partial u_m}{\partial \alpha_i} =
(-)^{n-m} e^{[i]}_{n-1-m}(\alpha)
\label{dudalpha}
\ee
Here $m=0,\ldots,n-1$, $i=1,\ldots, n$, and
$e^{[i]}_{m}(\alpha)$ is a symmetric polynomial of degree $m$ of the set of
$n-1$ variables consisting of $\alpha_j$ with $j\neq i$. See \cite{IM4, IM6} for more details
about the notations. Inverting (\ref{dudalpha}) we obtain:

\be
\frac{\partial \alpha_i}{\partial u_m} =
-\frac{\alpha_i^m}{\Delta_i}
\label{dalphadu}
\ee
Indeed, since $ \displaystyle{\sum_{m=0}^{n-1} (-)^{n-m} e^{[i]}_{n-1-m}(\alpha) x^m =
-\prod_{k\neq i}(x - \alpha_k)}$, we have:\footnote{
As an example of (\ref{dalphadu}),  we have for $n=2$:
$$
u_0 = \alpha_1\alpha_2,\ \ u_1 = -(\alpha_1+\alpha_2),
%\nn \\
$$ $$
\alpha_{1,2} = -\frac{1}{2}(u_1 \pm \sqrt{u_1^2 - 4u_0}),\ \
\Delta_1 = -\Delta_2 = \alpha_{12} = \mp \sqrt{u_1^2 - 4u_0},
%\nn \\
$$ $$
\frac{\partial\alpha_{1,2}}{\partial u_0} = \pm\frac{1}{\sqrt{u_1^2 - 4u_0}}
= -\frac{1}{\Delta_{1,2}}, \ \ \
\frac{\partial\alpha_{1,2}}{\partial u_1}
= -\frac{1}{2}\left(1 \pm\frac{u_1}{\sqrt{u_1^2 - 4u_0}}\right)
%= \pm\frac{\alpha_{1,2}}{\mp\Delta_{1,2}}
= -\frac{\alpha_{1,2}}{\Delta_{1,2}}
$$
}

\be
\sum_{m=0}^{n-1} (-)^{n-m} e^{[i]}_{n-1-m}(\alpha)\cdot
\left(-\frac{\alpha_j^m}{\Delta_j}\right)
= \frac{1}{\Delta_j} \prod_{k\neq i}(\alpha_j - \alpha_k) = \delta_{ij}.
\label{invert}
\ee

\section{Second derivatives of the prepotential: consistency
of eq.(\ref{derFT})}

   From (\ref{derFT}), (\ref{T=u})  and (\ref{tildeSvsS}) we obtain:

\be
\frac{\partial{\cal F}_{DV}}{\partial u_m} =
\frac{g}{m+1} res_\infty\ (x^{m+1}-\Lambda^{m+1}) dS_{DV}
%- ???\Lambda^{m+1}\sum_{i=1}^n S_i
= \frac{g}{m+1} \sum_{i=1}^n\tilde S_i(\alpha_i^{m+1}-\Lambda^{m+1})
%-???\Lambda^{m+1}\sum_{i=1}^n S_i
= \nn \\ =
\frac{1}{m+1}\left(g\sum_{i=1}^n S_i(\alpha_i^{m+1}-\Lambda^{m+1}) -
\frac{1}{2}\sum_{i,j,k=1}^n \eta_{i;jk}S_jS_k\alpha_i^{m+1} + \ldots\right)
%-???\Lambda^{m+1}\sum_{i=1}^n S_i
\label{prepder}
\ee
Substituting explicit expressions (\ref{eta}) for $\eta_{i;jk}$,
we obtain for the $S^2$-term:

\be
-\frac{1}{2(m+1)}\sum_{i,j,k=1}^n\eta_{i;jk}\alpha_i^{m+1} S_jS_k =
%\nn \\ =
-\frac{1}{2(m+1)}\sum_{i=1}^n\alpha_i^{m+1}
\sum_{j\neq i}
\left(\frac{S_i^2}{\alpha_{ij}^2\Delta_j} -
\frac{S_j^2}{\alpha_{ij}^2\Delta_i}\right) -
\nn\\ -
\frac{1}{m+1}\sum_{\stackrel{i,j}{j\neq i}}S_iS_j\left(\frac{2}{\alpha_{ij}^2} +
\frac{1}{\alpha_{ij}}\sum_{k\neq i,j}\frac{1}{\alpha_{ik}}\right)
\frac{\alpha_i^{m+1}}{\Delta_i} +
\frac{1}{2(m+1)}\sum_{i\neq j\neq k\neq i}S_jS_k\frac{\alpha_i^{m+1}}
{\alpha_{ij}\alpha_{ik}\Delta_i} = \nn \\
\stackrel{eq.(\ref{idem})}{=}\
-\frac{1}{2(m+1)}\sum_{\stackrel{i,j}{j\neq i}}\frac{S_i^2}{\alpha_{ij}^2}
\left(\frac{\alpha_i^{m+1} - \alpha_j^{m+1}}{\Delta_j}
\right) -
%\nn\\ -
\sum_{i<j} \frac{S_iS_j}{\alpha_{ij}}
\left(\frac{\alpha_i^{m}}{\Delta_i} - \frac{\alpha_j^{m}}{\Delta_j}
\right)
\label{quadra}
\ee
Since eq.(\ref{idem}) was used, (\ref{quadra}) is valid only for $m \leq n$.

Now we are ready to check the consistency of equation (\ref{prepder}),
i.e. the property of the matrix
$\partial^2{\cal F}_{DV}/\partial u_l\partial u_m$ being symmetric.
To the first order in $S$, the contribution is indeed symmetric
under an exchange of $l$ and $m$:

\be
\sum_{i=1}^n S_i\alpha_i^{m}\frac{\partial \alpha_i}{\partial u_l}
\ \stackrel{eq.(\ref{dalphadu})}{=}\
-\sum_{i=1}^n S_i\frac{\alpha_i^{l+m}}{\Delta_i}
\ee
Similarly, to the second order in $S_i$,  from the second term in the r.h.s. of
(\ref{quadra})  we find:

\be
-\frac{\partial}{\partial u_l}\
\sum_{i<j} \frac{S_iS_j}{\alpha_{ij}}
\left(\frac{\alpha_i^{m}}{\Delta_i} - \frac{\alpha_j^{m}}{\Delta_j}
\right) = \nn\\ =
\sum_{i<j} S_iS_j\left[
-\frac{1}{\alpha_{ij}^2}
\left(\frac{\alpha_i^l}{\Delta_i}-\frac{\alpha_j^l}{\Delta_j}\right)
\left(\frac{\alpha_i^m}{\Delta_i}-\frac{\alpha_j^m}{\Delta_j}\right)
+ \right.\nn \\ \left.
+ \frac{1}{\alpha_{ij}}\left(\frac{m\alpha_i^{l+m-1}}{\Delta_i^2} -
\frac{m\alpha_j^{l+m-1}}{\Delta_j^2}\right)
- \right.\nn \\ \left. -
\frac{\alpha_i^m}{\alpha_{ij}\Delta_i}\sum_{k\neq i}
\frac{1}{\alpha_{ik}}
\left(\frac{\alpha_i^l}{\Delta_i} - \frac{\alpha_k^l}{\Delta_k}\right)
%+ \right.\nn \\ \left.
+ \frac{\alpha_j^m}{\alpha_{ij}\Delta_j}\sum_{k\neq j}
\frac{1}{\alpha_{jk}}
\left(\frac{\alpha_j^l}{\Delta_j} - \frac{\alpha_k^l}{\Delta_k}\right)\right]
= \nn \\
\stackrel{eq.(\ref{ideem})}{=}\
\sum_{i<j} S_iS_j\left[
-\frac{1}{\alpha_{ij}^2}
\left(\frac{\alpha_i^l}{\Delta_i}-\frac{\alpha_j^l}{\Delta_j}\right)
\left(\frac{\alpha_i^m}{\Delta_i}-\frac{\alpha_j^m}{\Delta_j}\right)
+ \right.\nn \\ \left.
+ \frac{l+m}{\alpha_{ij}}\left(\frac{\alpha_i^{l+m-1}}{\Delta_i^2} -
\frac{\alpha_j^{l+m-1}}{\Delta_j^2}\right)
-\frac{2}{\alpha_{ij}}
\left(\frac{\alpha_i^{l+m}}{\Delta_i^2}\sum_{k\neq i}\frac{1}{\alpha_{ik}} -
\frac{\alpha_j^{l+m}}{\Delta_j^2}\sum_{k\neq j}\frac{1}{\alpha_{jk}}\right)
\right]
\label{symm2}
\ee
which is obviously symmetric under an exchange of $l$ and $m$.

As for the first term in (\ref{quadra}), it should first be transformed
with the help of (\ref{idemm}):

\be
-\frac{1}{2(m+1)}\sum_{\stackrel{i,j}{j\neq i}}\frac{S_i^2}{\alpha_{ij}^2}
\left(\frac{\alpha_i^{m+1} - \alpha_j^{m+1}}{\Delta_j}\right) =
\frac{1}{2}\sum_i \frac{S_i^2}{\Delta_i} \left(
\alpha_i^m\sum_{j\neq i}\frac{1}{\alpha_{ij}} \ -
\frac{m}{2}\alpha_i^{m-1}\right)
\ee
Now differentiation over $u_l$ gives:

\be
\frac{\partial}{\partial u_l} \left[\frac{1}{2}\sum_i \frac{S_i^2}{\Delta_i} \left(
\alpha_i^m\sum_{j\neq i}\frac{1}{\alpha_{ij}} \ -
\frac{m}{2}\alpha_i^{m-1}\right)
\right] = \nn \\ =
\frac{1}{2}\sum_i S_i^2 \left[
-\frac{m\alpha_i^{l+m-1}}{\Delta_i^2}\sum_{j\neq i}\frac{1}{\alpha_{ij}} +
\frac{m(m-1)}{2}\frac{\alpha_i^{l+m-2}}{\Delta_i^2} + \right.\nn \\ \left. +
\frac{1}{\Delta_i}\left(\alpha_i^m\sum_{j\neq i}\frac{1}{\alpha_{ij}} \ -
\frac{m}{2}\alpha_i^{m-1}\right)
\sum_{j\neq i}\frac{1}{\alpha_{ij}}\left(\frac{\alpha_i^l}{\Delta_i} -
\frac{\alpha_j^l}{\Delta_j}\right) + \right.\nn \\ \left. +
\frac{\alpha_i^m}{\Delta_i}\sum_{j\neq i}
\frac{1}{\alpha_{ij}^2}\left(\frac{\alpha_i^l}{\Delta_i} -
\frac{\alpha_j^l}{\Delta_j}\right)
\right]
\ee
and an application of (\ref{ideem}) to the second line
 and (\ref{idemm}) to the
third line finally provides a symmetric expression
(under an exchange of $l$ and $m$):

\be
\frac{1}{2}\sum_i \frac{S_i^2}{\Delta_i^2}\left[
-m\alpha_i^{l+m-1}\sum_{j\neq i}\frac{1}{\alpha_{ij}} +
\frac{m(m-1)}{2}\alpha_i^{l+m-2} + \right.\nn \\ \left. +
2\alpha_i^{l+m}\left(\sum_{j\neq i}\frac{1}{\alpha_{ij}}\right)^2 -
m\alpha_i^{l+m-1}\sum_{j\neq i}\frac{1}{\alpha_{ij}} -
l\alpha_i^{l+m-1}\sum_{j\neq i}\frac{1}{\alpha_{ij}} +
\frac{lm}{2}\alpha_i^{l+m-2}
- \right.\nn \\ \left.
-l\alpha_i^{l+m-1}\sum_{j\neq i}\frac{1}{\alpha_{ij}} +
\frac{l(l-1)}{2}\alpha_i^{l+m-2}
\right] = \nn \\ =
\sum_i \frac{S_i^2}{\Delta_i^2}\left[
\alpha_i^{l+m}\left(\sum_{j\neq i}\frac{1}{\alpha_{ij}}\right)^2
-(l+m)\alpha_i^{l+m-1}\sum_{j\neq i}\frac{1}{\alpha_{ij}} +
\frac{(l+m)^2-(l+m)}{4}\alpha_i^{l+m-2}
\right]
\ee

\section{Prepotential from its $T$-derivatives}

After consistency of the system (\ref{prepder}) is checked, one
can integrate these equations to obtain the prepotential.
Instead one can just check that the $u_m$-derivative of
(\ref{prepo}) is indeed equal to the r.h.s. of eq.(\ref{prepder}):

\be
\frac{\partial{\cal F}_{DV}}{\partial u_m} =
g\sum_{i=1}^n S_i \left(\frac{\partial W_{n+1}(\alpha_i)}{\partial u_m} -
\frac{\partial W_{n+1}(\Lambda)}{\partial u_m}\right) - \nn \\
-\frac{1}{4}\sum_{j<k} (S_j^2+S_k^2-4S_jS_k)\frac{1}{\alpha_{jk}}
\frac{\partial\alpha_{jk}}{\partial u_m} + \ldots
\ee
The check for the linear terms in $S_i$ is trivial: one should just take into
account that $\partial W_{n+1}(x)/\partial u_m = x^{m+1}/(m+1)$
and $ W_{n+1}'(\alpha_i) = P_n(\alpha_i) = 0$. As for the quadratic terms in $S_i$,
comparison with (\ref{prepder}) implies that

\be
\frac{1}{\alpha_{jk}}\frac{\partial\alpha_{jk}}{\partial u_m} =
-\frac{1}{m+1}\sum_{i=1}^n \eta_{i;jk}\alpha_i^{m+1} = \nn\\
= -\frac{1}{m+1}
\left(\eta_{j;jk}\alpha_j^{m+1} + \eta_{k;jk}\alpha_k^{m+1} +
\sum_{i\neq j,k}^n \eta_{i;jk}\alpha_i^{m+1}\right),
\ \ {\rm for}\ j<k;
\nn \\
\frac{1}{4}\sum_{k\neq j}
\frac{1}{\alpha_{jk}}\frac{\partial\alpha_{jk}}{\partial u_m}
= \frac{1}{2}\left(\eta_{j;jj}\alpha_j^{m+1} +
\sum_{i\neq j}\eta_{i;jj}\alpha_i^{m+1}\right)
%\nn\\
%\sum_i \eta_{i,jj}\alpha_i^{m+1} = -2(m+1)\sum_{k\neq j}
%\frac{1}{\alpha_{jk}}\frac{\partial\alpha_{jk}}{\partial u_m} =
%-\frac{1}{2}\sum_{k\neq j} \sum_i\eta_{i;jk}\alpha_i^{m+1}
\ee
Consistency of
these two relations is ensured by the second sum rule (\ref{sumruleseta}).
Relations can be checked with the help of identities (\ref{idem})
and (\ref{idemm}) respectively, after explicit expressions
are substituted for $\eta_{i;jk}$ from (\ref{eta}) and for
$\partial\alpha_{jk}/\partial u_m = -\alpha_j^m/\Delta_j +
\alpha_k^m/\Delta_k$.

These calculations are straightforwardly generalized to the higher order
contributions ${\cal F}_p$ to the prepotential in $S_i$: one should just
consider next terms in the expansion (\ref{dSexpansion}) and
repeat all the steps of the above procedure.

\section{${\cal L}_{-1}$ and ${\cal L}_{0}$-constraints on ${\cal F}_{DV}$
\label{WId}}

The usual way to derive Ward identities (see, for example, \cite{loop})
is to shift integration variables without changing the integral.
This shift changes the shape of the {\it integrand},
which is equivalent to a certain change of its parameters (a shift along
the moduli space), so that invariance of the integral provides a
differential equation for it. In our present case this reasoning can be
applied to the integrals (\ref{moduli}), (\ref{derFS}) and (\ref{derFT}).

The freedom to shift the coordinate $x$ on the spectral surface
  by $x \rightarrow x+\epsilon$ is equivalent to

\be
\alpha_i \rightarrow\alpha_i + \epsilon,\ \ \
u_{k-1} \rightarrow u_{k-1} - \epsilon ku_k,\ \ \
\Lambda \rightarrow \Lambda + \epsilon; \nn\\
S_i \rightarrow S_i, \ \ \
\frac{\partial{\cal F}_{DV}}{\partial S_i} \rightarrow
\frac{\partial{\cal F}_{DV}}{\partial S_i} \;\;.
\ee
This implies a constraint on ${\cal F}_{DV}(S|T)$:

\be
{\cal L}_{-1}{\cal F}_{DV}(S|T) =
\frac{\partial}{\partial\Lambda}{\cal F}_{DV}(S|T)
\label{Vir-1}
\ee
where

\be
{\cal L}_{-1} = \sum_{k=1}^n ku_k\frac{\partial}{\partial u_{k-1}}
= gn\frac{\partial}{\partial T_n} +
\sum_{k=1}^{n-1} kT_{k+1}\frac{\partial}{\partial T_k}
\ee

The CIV-DV prepotential indeed satisfies this constraint, for which
the presence of the term $W_{n+1}(\Lambda)\sum_{i=1}^n S_i$ is
essential however. Note that

\be
\left({\cal L}_{-1} - \frac{\partial}{\partial\Lambda}\right)
W_{n+1}(\Lambda) =
\left({\cal L}_{-1} - \frac{\partial}{\partial\Lambda}\right)
W_{n+1}(\alpha_i) =
-u_0 = -\frac{1}{g}T_1.
\ee

Since $u_n=1=const$, the derivative with respect to $T_n = gu_{n-1}$
appears in the constraint (\ref{Vir-1})
with a moduli-independent coefficient.
This implies that the dependence of the prepotential on $T_n$
is easy to restore once its dependence on all other moduli is
known. However, since the constraint is inhomogeneous in $T$'s,
the $T_n$-dependence can not be simply ignored 
 -- the constraint (\ref{Vir-1}) 
does not commute with
the $T$-derivatives and using it to eliminate the $T_n$-dependence
changes, say, the form of the WDVV equations.

Similarly, rescaling of $x$, $x \lambda\longrightarrow x$,
equivalent to

\be
\alpha_i \rightarrow\lambda\alpha_i,\ \ \
u_{k} \rightarrow \lambda^{n-k} u_{k},\ \ \,
\Lambda \rightarrow \lambda\Lambda; \nn \\
S_i \rightarrow \lambda^{n+1}S_i, \ \ \
\frac{\partial{\cal F}_{DV}}{\partial S_i} \rightarrow
\lambda^{n+1}\frac{\partial{\cal F}_{DV}}{\partial S_i},
\ee
provides the ${\cal L}_{0}$-constraint on ${\cal F}_{DV}$:\footnote{
Of course, the prepotential (\ref{prepo}) satisfies this and, what is
 more, each individual term does. For example,
for
$$
S_iW_{n+1}(\Lambda) = S_i \sum_{k=0}^n \frac{u_k}{k+1}\Lambda^{k+1}
$$
we have
$$
\sum_{k=0}^n\frac{n-k}{k+1}u_k\Lambda^{k+1} = -\Lambda W_{n+1}'(\Lambda) +
(n+1)(2W_{n+1}(\Lambda) - W_{n+1}(\Lambda)).
$$
}

\be
{\cal L}_{0}{\cal F}_{DV}(S|T) =
\frac{\partial}{\partial\log\Lambda}{\cal F}_{DV}(S|T) -
(n+1)\left(2 - \sum_i S_i\frac{\partial}{\partial S_i}\right)
{\cal F}_{DV}(S|T)
\label{Vir0}
\ee
where

\be
{\cal L}_{0} = -\sum_{k=0}^{n-1} (n-k)u_k\frac{\partial}{\partial u_{k}}
= \sum_{k=1}^{n} (n+1-k)T_{k}\frac{\partial}{\partial T_k}
\ee
From eq.(\ref{derFT}) we obtain

\be
{\cal L}_{0}{\cal F}_{DV}(S|T) = - g \sum_{i=1}^n\left(
\sum_{k=0}^{n-1} \frac{n-k}{k+1} u_k
(\alpha_i^{k+1} - \Lambda^{k+1})\tilde S_i
\right)
\label{summ}
\ee
At the special Seiberg-Witten ($N=2$ supersymmetric) point, where
$Y^2(x) = g^2 \left( P_n(x)^2 - \Lambda_{N=2}^{2n} \right)$, i.e.

\be
\tilde S_i = -\frac{g\Lambda_{N=2}^{2n}}{2\Delta_i}
\ee
and

\be
\sum_i \frac{\alpha_i^{k+1}}{\Delta_i} =
\ res_\infty \frac{x^{k+1}dx}{P_n(x)}
= \delta_{k,n-2} - u_{n-1}\delta_{k,n-1}
\ee
(the sum is non-vanishing for $k \geq n$ as well, but such quantities
do not appear in (\ref{summ})),
one obtains from (\ref{Vir0}) Matone's identity
\cite{Mat,DP} (for any $n$):\footnote{
Note that, for the sake of simplicity, we have changed our notation
 from the usual one:
our $u_k$ is conventional $u_{n-k}$ and, in particular,
our $u_{n-2}$ is conventional $u_2$.}

\be
-(n+1)\left(2 - \sum_i S_i\frac{\partial}{\partial S_i}\right) {\cal F}_{DV} =
 g^{2} \Lambda_{N=2}^{2n}\left(\frac{u_{n-2}}{n-1} - \frac{u_{n-1}^2}{2n}\right) \;\;.
\ee
The term $\displaystyle{W_{n+1}(\Lambda)\sum_{i=1}^n S_i}$ does not
 contribute, because at the SW point
$\displaystyle{\sum_{i=1}^n S_i = 0}$.

\section{Third derivatives of the prepotential and WDVV equations}

Generic proof of the WDVV equations \cite{MMM},

\be
\check{\cal F}_I \check{\cal F}_J^{-1} \check{\cal F}_K  =
\check{\cal F}_K \check{\cal F}_J^{-1} \check{\cal F}_I,
\label{WDVV}
\ee
consists of deriving residue formula for the
third derivatives,

\be
(\check{\cal F}_I)_{JK} = \frac{\partial {\cal F}}
{\partial\mu_I\partial\mu_J\partial\mu_K} =
\sum_{\stackrel{dS=0\ or}{ddS/dS=0}}
res\
\frac{dW_IdW_JdW_K}{ddS} \;\;\;,
\ee
and considering the closed "algebra" of one-differentials,

\be
dW_IdW_J = C_{IJ}^K dW_K (\eta_L dW_L) \ mod(ddS/dS) \;\;.
\ee
  Here  $\{\mu_I\} = \{S_i,T_i\}$ is the set of {\it flat} moduli,
$dW_I$ are the corresponding {\it canonical} one-differentials,
in DV model, and the two-differential
$ddS \sim dS_{DV}(x)d\log Y(x) = dY(x)dx$.
See \cite{MMM,IM4} for further details.
 The proof of such kind for the CIV-DV prepotential
was discussed in \cite{IM4} (for {\it regularized} DV
model) and \cite{Chetal}, and we do not repeat it here.
Instead we check the validity of the WDVV equations to
the leading order in the expansion in $S$ (or $g^{-1}$)
with the help of explicit formulas for ${\cal F}_{DV}(S,T)$
from the previous sections.

Because of the existence of two types of moduli, $S_i$ and $T_i$,
the $2n\times 2n$ matrices of the prepotential third derivatives
naturally possess the block form \footnote{In accordance with (\ref{T=u})
the $T_m$ derivatives are actually taken with respect to $g u_{m-1}$.}:

\be
\check{\cal F}_{S_i} =
\left(\begin{array}{cc}
\frac{1}{2S_i}\check\Pi_i & \check{\cal B}^{(i)} \\ & \\
\tilde{\check{\cal B}^{(i)}} & g\check{\cal C}^{(i)}
\end{array}\right) =
\left(\begin{array}{cc}
\frac{1}{2S_i}\delta_{ij}\delta_{ik} & \check{\cal B}^{(i)}_{jl} \\
& \\
\check{\cal B}^{(i)}_{mk} & g\check{\cal C}^{(i)}_{ml}
\end{array}\right)
\ee

\be
\check{\cal F}_{u_r} =
\left(\begin{array}{cc}
\check{\cal B}_r  & g\check{\cal C}_r \\
 g\tilde{\check{\cal C}_r}  & 0
\end{array}\right) =
\left(\begin{array}{cc}
I & 0 \\ 0  & g\tilde{\check{\cal C}_r}
\end{array}\right)\left(\begin{array}{cc}
\check{\cal B}_r  & I \\
 I  & 0
\end{array}\right)\left(\begin{array}{cc}
I & 0 \\ 0  & g\check{\cal C}_r
\end{array}\right) =
\nn \\ =
\left(\begin{array}{cc}
(\check{\cal B}_r)_{jk} & g(\check{\cal C}_r)_{jl} \\
g(\check{\cal C}_r)_{mk} & 0
\end{array}\right)
\label{Fu}
\ee
Tilde denotes a transposition of matrices.
We neglect here all the contributions with higher powers of $g^{-1}$
(i.e. with higher powers of $S$) in each block. If not the singular
$1/S$ item, these would be the values of matrices for $S=0$.
The entries are:

\be
(\check{\cal B}_m)_{ij} = \check{\cal B}^{(i)}_{jm}
= \frac{1}{2}\delta_{ij}\sum_{k\neq i}
\left(\frac{\alpha_i^m}{\Delta_i} - \frac{\alpha_k^m}{\Delta_k}\right)
\frac{1}{\alpha_{ik}} - \frac{1-\delta_{ij}}{\alpha_{ij}}
\left(\frac{\alpha_i^m}{\Delta_i} - \frac{\alpha_j^m}{\Delta_j}\right),
\nn \\
(\check{\cal C}_m)_{il} = \check{\cal C}^{(i)}_{lm}
=- \frac{\alpha_i^{l+m}}{\Delta_i}
\label{BandC}
\ee
Matrices $\check{\cal B}_m$ and $\check{\cal C}^{(i)}$ are symmetric,
but $\check{\cal B}^{(i)}$ and $\check{\cal C}_m$ are not.
The matrix $\check{\cal F}_{u_r}$ (in variance with $\check{\cal F}_{S_i}$)
can be easily inverted:

\be
\check{\cal F}_{u_r}^{-1} =
\left(\begin{array}{cc}
I & 0 \\ 0  & \frac{1}{g}\tilde{\check{\cal C}_r^{-1}}
\end{array}\right)\left(\begin{array}{cc}
0  & I \\
 I  & -\check{\cal B}_r
\end{array}\right)\left(\begin{array}{cc}
I & 0 \\ 0  & \frac{1}{g}{\check{\cal C}_r^{-1}}
\end{array}\right) =
\nn \\ =
\left(\begin{array}{cc}
0  & \frac{1}{g}\tilde{\check{\cal C}}_r^{-1} \\
\frac{1}{g}{\check{\cal C}}_r^{-1}  &
-\frac{1}{g^2}{\check{\cal C}}_r^{-1} \check{{\cal B}}_r
\tilde{\check{\cal C}}_r^{-1}
\end{array}\right)
\label{Fuinv}
\ee
Matrix $\check {\cal C}_r$ in its turn is decomposed into diagonal and
$r$-independent matrices:

\be
\check {\cal C}_r = \check {\cal A}^r\check{\cal C};\ \ \
\check{\cal A} = \ diag(\alpha_i), \ i.e.\
\check{\cal A}^r_{ij} = \delta_{ij}\alpha_i^r;\ \ \
\check{\cal C}_{im} = - \frac{\alpha_i^{m}}{\Delta_i}
\label{Crdec}
\ee
Both are easy to invert (use (\ref{invert}) in the case of $\check{\cal C}$):

\be
\check {\cal C}_r^{-1} = \check{\cal C}^{-1} \check {\cal A}^{-r};\ \ \
\check{\cal A}^{-r}_{ij} = \delta_{ij}\alpha_i^{-r};\ \ \
(\check{\cal C}^{-1})_{mi} = (-)^{n-m} e^{[i]}_{n-1-m}(\alpha).
\label{C-1}
\ee

\section{WDVV equations, explicit check (in the leading order only)}

For illustrative purposes we use the results of the previous section
to check explicitly the WDVV equations (\ref{WDVV}) to the leading
order in power series in $S$.
As we saw in the previous section, matrices
$\check{\cal F}_{u_r}(S=0)$ are especially easy to invert
 and we will exploit this.
Actually, it is enough to check (\ref{WDVV}) for any particular
 $J$ but for all $I$ and $K$. Equations for other $J$ follow
 automatically. See \cite{MMM}. Indices $I,K$ still correspond to
either $S_k$ or $T_k$ however.

The simplest case is when all the three moduli $\mu_{I,J,K}$
in (\ref{WDVV}) are Whitham times $T_{r+1} = gu_r$:

\be
\check{\cal F}_{u_{r_1}} \check{\cal F}_{u_{r_2}}^{-1}
\check{\cal F}_{u_{r_3}} =
\check{\cal F}_{u_{r_3}} \check{\cal F}_{u_{r_2}}^{-1}
\check{\cal F}_{u_{r_1}}
\label{WDVVT}
\ee
In the leading order, we can $S_i=0$ in (\ref{WDVVT}),
while some entries of $\check{\cal F}_{S_j}$ are singular when
$S=0$. This means that one can use (\ref{Fu}) and (\ref{Fuinv})
in (\ref{WDVVT}), which becomes a condition for
the matrix being symmetric   

\be
\left(\begin{array}{cc}
\check{\cal B}_{r_1}  & g\check{\cal C}_{r_1} \\
 g\tilde{\check{\cal C}_{r_1}}  & 0
\end{array}\right)
\left(\begin{array}{cc}
0  & \frac{1}{g}\tilde{\check{\cal C}_{r_2}^{-1}} \\
\frac{1}{g}\check{\cal C}_{r_2}^{-1}  &
-\frac{1}{g^2}\check{\cal C}_{r_2}^{-1}\check{\cal B}_{r_2}
\tilde{\check{\cal C}_{r_2}^{-1}}
\end{array}\right)
\left(\begin{array}{cc}
\check{\cal B}_{r_3}  & g\check{\cal C}_{r_3} \\
 g\tilde{\check{\cal C}_{r_3}}  & 0
\end{array}\right) = \nn \\
= \left(\begin{array}{cc}
\check{\cal B}_{r_1}  & g\check{\cal C}_{r_1} \\
 g\tilde{\check{\cal C}_{r_1}}  & 0
\end{array}\right)
\left(\begin{array}{ccc}
\tilde{\check{\cal C}_{r_2}^{-1}}\tilde{\check{\cal C}_{r_3}} &  \ & 0 \\
\frac{1}{g}\check{\cal C}_{r_2}^{-1}(\check{\cal B}_{r_3} -
\check{\cal B}_{r_2}
\tilde{\check{\cal C}_{r_2}^{-1}}\tilde{\check{\cal C}_{r_3}}) &  \ &
\check{\cal C}_{r_2}^{-1}\check{\cal C}_{r_3}
\end{array}\right) =
\nn \\ =
\left(\begin{array}{ccc}
\check{\cal B}_{r_1}\tilde{\check{\cal C}_{r_2}^{-1}}
\tilde{\check{\cal C}_{r_3}} +
\check{\cal C}_{r_1} \check{\cal C}_{r_2}^{-1} \check{\cal B}_{r_3} -
\check{\cal C}_{r_1} \check{\cal C}_{r_2}^{-1} \check{\cal B}_{r_2}
\tilde{\check{\cal C}_{r_2}^{-1}}
\tilde{\check{\cal C}_{r_3}} & \ &
g\check{\cal C}_{r_1}\check{\cal C}_{r_2}^{-1}\check{\cal C}_{r_3} \\
g\tilde{\check{\cal C}_{r_1}}
\tilde{\check{\cal C}_{r_2}^{-1}}\tilde{\check{\cal C}_{r_3}} &  \ & 0
\end{array}\right)  \;\;.
\ee
We  need to check that

\be
\check{\cal C}_{r_1} \check{\cal C}_{r_2}^{-1} \check{\cal C}_{r_3} =
\check{\cal C}_{r_3} \check{\cal C}_{r_2}^{-1} \check{\cal C}_{r_1}
\label{r123_1}
\ee
 as well as

\be
\check{\cal B}_{r_1}\tilde{\check{\cal C}_{r_2}^{-1}}
\tilde{\check{\cal C}_{r_3}} +
\check{\cal C}_{r_1} \check{\cal C}_{r_2}^{-1} \check{\cal B}_{r_3} -
\check{\cal C}_{r_1} \check{\cal C}_{r_2}^{-1} \check{\cal B}_{r_2}
\tilde{\check{\cal C}_{r_2}^{-1}}
\tilde{\check{\cal C}_{r_3}} = \nn \\ =
\check{\cal C}_{r_3} \check{\cal C}_{r_2}^{-1} \check{\cal B}_{r_1} +
\check{\cal B}_{r_3}\tilde{\check{\cal C}_{r_2}^{-1}}
\tilde{\check{\cal C}_{r_1}} -
\check{\cal C}_{r_3} \check{\cal C}_{r_2}^{-1} \check{\cal B}_{r_2}
\tilde{\check{\cal C}_{r_2}^{-1}}
\tilde{\check{\cal C}_{r_1}}
\label{r123_2}
\ee
This is straightforward (with the aid of explicit formulas from the previous
section). Indeed, from (\ref{Crdec})

\be
\check{\cal C}_{r_1} \check{\cal C}_{r_2}^{-1} =
\tilde{\check{\cal C}_{r_2}^{-1}}
\tilde{\check{\cal C}_{r_1}} =
\check{\cal A}^{r_1-r_2} \;\;,
\label{CC-1}
\ee
so that the entries in (\ref{r123_1}) are obviously identical:

\be
\check{\cal C}_{r_1} \check{\cal C}_{r_2}^{-1} \check{\cal C}_{r_3} =
\check{\cal A}^{r_1-r_2+r_3}\check{\cal C} =
\check{\cal C}_{r_3} \check{\cal C}_{r_2}^{-1} \check{\cal C}_{r_1} \;\;.
\ee
 As for (\ref{r123_2}), it reduces to

\be
\check{\cal B}_{r_1}\check{\cal A}^{r_3-r_2} +
\check{\cal A}^{r_1-r_2}\check{\cal B}_{r_3} -
\check{\cal A}^{r_1-r_2}\check{\cal B}_{r_2}\check{\cal A}^{r_3-r_2} =
\nn \\ =
\check{\cal A}^{r_3-r_2}\check{\cal B}_{r_1} +
\check{\cal B}_{r_3}\check{\cal A}^{r_1-r_2} -
\check{\cal A}^{r_3-r_2}\check{\cal B}_{r_2}\check{\cal A}^{r_1-r_2}
\ee
which is easily checked by substituting explicit
expressions for $\check{\cal B}$ from (\ref{BandC}).

More sophisticated is the check of another subset of eqs.(\ref{WDVV}),

\be
\check{\cal F}_{S_{i}} \check{\cal F}_{u_{r_2}}^{-1}
\check{\cal F}_{u_{r_3}} =
\check{\cal F}_{u_{r_3}} \check{\cal F}_{u_{r_2}}^{-1}
\check{\cal F}_{S_{i}}
\label{WDVVi23}
\ee
At the l.h.s. we have in our approximation:

\be
\left(\begin{array}{cc}
\frac{1}{2S_i}\check\Pi_i & \check{\cal B}^{(i)} \\ & \\
\tilde{\check{\cal B}^{(i)}} & g\check{\cal C}^{(i)}
\end{array}\right)
\left(\begin{array}{cc}
0  & \frac{1}{g}\tilde{\check{\cal C}_{r_2}^{-1}} \\
\frac{1}{g}\check{\cal C}_{r_2}^{-1}  &
-\frac{1}{g^2}\check{\cal C}_{r_2}^{-1}\check{\cal B}_{r_2}
\tilde{\check{\cal C}_{r_2}^{-1}}
\end{array}\right)
\left(\begin{array}{cc}
\check{\cal B}_{r_3}  & g\check{\cal C}_{r_3} \\
 g\tilde{\check{\cal C}_{r_3}}  &  0
\end{array}\right) = \nn \\
= \left(\begin{array}{cc}
 \frac{1}{2S_i}\check\Pi_i & \check{\cal B}^{(i)} \\ & \\
\tilde{\check{\cal B}^{(i)}} & g\check{\cal C}^{(i)}
\end{array}\right)
\left(\begin{array}{ccc}
\tilde{\check{\cal C}_{r_2}^{-1}}\tilde{\check{\cal C}_{r_3}} &
\ & 0 \\
\frac{1}{g}\check{\cal C}_{r_2}^{-1}(\check{\cal B}_{r_3} -
\check{\cal B}_{r_2}
\tilde{\check{\cal C}_{r_2}^{-1}}\tilde{\check{\cal C}_{r_3}}) &  \ &
\check{\cal C}_{r_2}^{-1}\check{\cal C}_{r_3}
\end{array}\right) = \nn \\ =
\left(\begin{array}{ccc}
 \frac{1}{2S_i}\check\Pi_i\tilde{\check{\cal C}_{r_2}^{-1}}
\tilde{\check{\cal C}_{r_3}} + O(g^{-1}) & \ &
\check{\cal B}^{(i)}\check{\cal C}_{r_2}^{-1}\check{\cal C}_{r_3}
% -\frac{1}{2S_i}\check\Pi_i \varepsilon
\\ & & \\
\tilde{\check{\cal B}^{(i)}}
\tilde{\check{\cal C}_{r_2}^{-1}}\tilde{\check{\cal C}_{r_3}} +
\check{\cal C}^{(i)}\check{\cal C}_{r_2}^{-1}
(\check{\cal B}_{r_3} - \check{\cal B}_{r_2}
\tilde{\check{\cal C}_{r_2}^{-1}}\tilde{\check{\cal C}_{r_3}})
&  \ & g\check{\cal C}^{(i)}\check{\cal C}_{r_2}^{-1}\check{\cal C}_{r_3}
\end{array}\right)
\label{prom}
\ee
WDVV equations do not really imply, however, that the entire
matrix at the r.h.s. of (\ref{prom}) is symmetric. This is because
some entires of $\check{\cal F}_{S_{i}}$ are singular at $S_i=0$,
and therefore some linear-in-$S$ contributions to
$\check{\cal F}_{u_{r_2}}$ and $\check{\cal F}_{u_{r_3}}$
(in particular, those denoted by zeroes in (\ref{prom}))
will contribute even in the limit $S=0$.
It is easy to see that such contributions can arise at the
upper-right corner of (\ref{prom}).

Thus one should check that the matrices at the upper-left
and lower-right corners are symmetric only. The first one is,

\be
\check\Pi_i \tilde{\check{\cal C}_{r_2}^{-1}}
\tilde{\check{\cal C}_{r_3}} =
\check\Pi_i \check{\cal A}^{r_3-r_2}
\ee
This is trivially symmetric under an interchange of $j$ and $l$:

\be
(\check\Pi_i \check{\cal A}^{r_3-r_2})_{jl} =
\delta_{ij}\delta_{il}\alpha_i^{r_3-r_2}
\ee
Whether the second matrix 

\be
\check{\cal C}^{(i)}\check{\cal C}_{r_2}^{-1} \check{\cal C}_{r_3}
\ee
is symmetric is a little less trivial to check, since now (\ref{CC-1}) is unapplicable
and one should use explicit expression
(\ref{C-1}) for $\check{\cal C}^{-1}$:

\be
(\check{\cal C}^{(i)}\check{\cal C}_{r_2}^{-1} \check{\cal C}_{r_3})_{km} =
\sum_l (\check{\cal C}^{(i)})_{kl}
(\check{\cal C}_{r_2}^{-1} \check{\cal C}_{r_3})_{lm} =
\sum_l (\check{\cal C}^{(i)})_{kl}
(\check{\cal C}^{-1}\check{\cal A}^{r_3-r_2} \check{\cal C})_{lm} =
\nn \\ =
\sum_{l,j} \frac{\alpha_i^{k+l}}{\Delta_i}e^{[j]}_{n-1-l}(\alpha)(-)^l
\alpha_j^{r_3-r_2}\frac{\alpha_j^m}{\Delta_j}
\label{cinvapp}
\ee
Summing over $l$ gives $\delta_{ij}\Delta_i$ and the entire expression
becomes equal to

\be
\frac{\alpha_i^{k+m+r_3-r_2}}{\Delta_i}
\ee
which is obviously symmetric under an interchange of $k$ and $m$.

As already mentioned, the remaining condition
for the matrix (\ref{prom}) being symmetric reads

\be
\check{\cal B}^{(i)}\check{\cal C}_{r_2}^{-1} \check{\cal C}_{r_3} -
\check{\cal C}_{r_3}\check{\cal C}_{r_2}^{-1} \check{\cal B}^{(i)}
+ \left(\check{\cal C}_{r_3}\check{\cal C}_{r_2}^{-1}
\check{\cal B}_{r_2} - \check{\cal B}_{r_3}  \right)
\tilde{\check{\cal C}_{r_2}^{-1}}\check{\cal C}^{(i)}
%= \nn \\
= - \frac{1}{2S_i}\check\Pi_i \check\varepsilon \;\;,
\label{promo}
\ee
 and this
depends on the $S$-linear contributions
to the upper-right corner of $\check{\cal F}_{u_{r_2}}^{-1}
\check{\cal F}_{u_{r_3}}$, denoted by $\check\varepsilon$
at the r.h.s. of (\ref{promo}).
Manipulations similar to (\ref{cinvapp})
transform the matrix element $jm$ of the l.h.s. of (\ref{promo}) to

\be
- \frac{1}{2}  \delta_{ij} \sum_{k\neq i}\frac{\alpha_k^{r_2}}{\alpha_{ik}\Delta_k}
(\alpha_k^{r_3-r_2} - \alpha_i^{r_3-r_2})
(\alpha_k^{m-r_2} - \alpha_i^{m-r_2}) \;\;.
\label{ntsum}
\ee
If the sum rule (\ref{ideem}) could be applied to this sum over $k$,
one would immediately conclude that it vanishes. However, one of the
contributions to (\ref{ntsum}) involves $\alpha_k^{m+r_3-r_2}$,
 whose exponent is $m + r_3 - r_2$. This last factor can be either negative or
exceed $n-1$. Whenever this happens, eq.(\ref{ideem}) is violated
by contributions from residues at zero or infinity, i.e. the
l.h.s. of (\ref{promo}) does not vanish and in this case the r.h.s.
should not vanish as well. Since the proof of this statement
takes us beyond the leading approximation, we do not go into
further details here.

The last subset of eqs.(\ref{WDVV}),

\be
\check{\cal F}_{S_{i}} \check{\cal F}_{u_{r_2}}^{-1} \check{\cal F}_{S_{k}}
= \check{\cal F}_{S_{k}} \check{\cal F}_{u_{r_2}}^{-1}
\check{\cal F}_{S_{i}}
\ee
can be analyzed in a similar way.

As already mentioned, the WDVV
identities with any $\check{\cal F}_{u_{S_j}}^{-1}$
standing in place of $\check{\cal F}_{u_{r_2}}^{-1}$
are not independent and do not require a separate validation.
See \cite{MMM}.

\section{Acknowledgements}
We are indebted for useful comments to A.Dymarsky,
H. Kanno, A.Mironov and V.Pestun.
A.M. acknowledges the hospitality
of the Osaka City University during his stay in Japan.
Our work is partly supported by the 
Grant-in-Aid for Scientific Research (14540284) from the
Ministry of Education, Science and Culture, Japan (H.I.),
and by the Russian President's grant 00-15-99296, by RFBR-01-02-17488,
INTAS 00-561 and by Volkswagen-Stiftung (A.M.).


\begin{thebibliography}{12}

\bibitem{CIV}
F.Cachazo, K.A.Intriligator and C.Vafa,
{\it Nucl.Phys.} {\bf B603} (2001) 3, hep-th/0103067; \\
F.Cachazo and C.Vafa, hep-th/0206017.

\bibitem{DV}
R.Dijkgraaf and C.Vafa,
{\it Nucl.Phys.} {\bf B644} (2002) 3-20, hep-th/0206255;
{\it Nucl.Phys.} {\bf B644} (2002) 21-39, hep-th/0207106;
hep-th/0208048.

\bibitem{followup} 
L.Chekhov and A.Mironov, hep-th/0209085; \\
N.Dorey, T.J.Hollowood, S.Prem Kumar and A.Sinkovics, 
hep-th/0209089; hep-th/0209099; \\
M.Aganagic and C.Vafa, hep-th/0209138;\\
F.Ferrari, hep-th/0210135; hep-th/0211069;\\
H.Fuji and Y.Ookouchi, hep-th/0210148; \\
D.Berenstein, hep-th/0210183; \\
R.Dijkgraaf, S.Gukov, V.Kazakov and C.Vafa,
hep-th/0210238; \\
N.Dorey, T.J.Hollowood and S.Prem Kumar, 
hep-th/0210239; \\
A.Gorsky, hep-th/0210281; \\
R.Argurio, V.L.Campos, G.Ferretti and R.Heise, hep-th/0210291; \\
J.McGreevy, hep-th/0211009;\\
R. Dijkgraaf, M.T. Grisaru, C.S. Lam, C. Vafa and D. Zanon,
hep-th/0211017;\\
H.Suzuki, hep-th/0211052; \\
I.Bena and R.Roiban, hep-th/0211075; \\
Y.Demasure and R.A.Janik, hep-th/0211082;\\
M.Aganagic, A.Klemm, M.Marino and C.Vafa, hep-th/0211098;\\
R.Gopakumar, hep-th/0211100; \\
S.Naculich, H.Schnitzer and N.Wyllard, hep-th/0211123; hep-th/0211254; \\
%F.Cachazo, M.R.Douglas, N.Seiberg and E.Witten, hep-th/0211170; \\
R.Dijkgraaf, A.Neitzke and C.Vafa, hep-th/0211194;\\
Y.Tachikawa, hep-th/0211189; hep-th/0211274;\\
B.Feng, hep-th/0211202; hep-th/0212010; hep-th/0212274;\\
B.Feng and Y.-H. He, hep-th/0211234;\\
A.Klemm, M.Marino and S.Theisen, hep-th/0211216; \\
V.Kazakov and A.Marshakov, hep-th/0211236; \\
R.Dijkgraaf, A.Sinkovics and M.Tmerhan, hep-th/0211241;\\
R.Argurio, V.L.Campos, G.Ferretti and R.Heise, hep-th/0211249;\\
H.Ita, H.Nieder, Y.Oz, hep-th/0211261; \\
I.Bena, R.Roiban and R.Tatar, hep-th/0211271; \\
Y.Ookouchi, hep-th/0211287; \\
S.K.Ashok, R.Corrado, N.Halmagyi, K.D.Kennaway and C.Romelsberger, hep-th/0211291;\\
K. Ohta, hep-th/0212025; \\
T. J. Hollowood, hep-th/0212065; \\
R.A. Janik and N.A. Obers, Phys.Lett.B553:309-316,2003, hep-th/0212069; \\
S. Seki, hep-th/0212079; \\
V. Balasubramanian, et al., hep-th/0212082; \\
I. Bena, et al., hep-th/0212083; \\
C. Hofman, hep-th/0212095; \\
H. Suzuki, hep-th/0212121; \\
M. Tierz, hep-th/0212128; \\
Y. Demasure and R. A. Janik, hep-th/0212212; \\
N. Seiberg, hep-th/0212225; \\
C. Ahn and S. Nam, hep-th/0212231; \\
A. Iqbal and A. Kashani-Poor, hep-th/0212279; \\
E. D'Hoker, I.Krichever, D.H. Phong,  hep-th/0212313; \\ 
R. Brustein, et al., hep-th/0212344; \\     
F. Cachazo, N.Seiberg and E.Witten, hep-th/0301006;\\
C. Ahn, hep-th/0301011

\bibitem{SW}
N.Seiberg and E.Witten, {\it Nucl.Phys.} {\bf B426}
(1994) 19 ( Erratum-ibid. {\bf B430} (1994) 485), hep-th/9407087;
{\it ibidem}, {\bf B431} (1994) 484, hep-th/9408099; hep-th/9607163; \\
A.Klemm, W.Lerche, S.Theisen and S.Yankielowicz,
{\it Phys.Lett.} {\bf B344} (1995) 169, hep-th/9411048; \\
P.Argyres and A.Farragi, {\it Phys.Rev.Lett.} {\bf 74} (1995) 3931,
hep-th/9411057; \\
A.Hanany and Y.Oz, {\it Nucl.Phys.} {\bf B452} (1995) 283,
hep-th/9505074; \\
N.Seiberg, {\it Phys.Lett.} {\it B388} (1996) 753-760.

\bibitem{intfollowup} 
A.Gorsky, I.Krichever, A.Marshakov, A.Mironov and A.Morozov,
{\it Phys.Lett.} {\bf B355} (1995) 466, hep-th/9505035; \\
E.Martinec and N.Warner, {\it Nucl.Phys.} {\bf B459} (1996) 97-112,
hep-th/9509161;  \\
T.Nakatsu and K.Takasaki, {\it Mod.Phys.Lett.} {\bf A11} (1996) 157-168,
hep-th/9509162; \\
R.Donagi and E.Witten, {\it Nucl.Phys.} {\bf B460} (1996) 299, hep-th/9510101;\\
T.Eguchi and S.Yang, {\it Mod.Phys.Lett.} {\bf A11} (1996) 131-138,
hep-th/9510183; \\
H.Itoyama and A.Morozov, {\it Nucl.Phys.} {\bf B477} (1996)
855, hep-th/9511126; hep-th/9601168.

\bibitem{IM4} H.Itoyama and A.Morozov, hep-th/0211245.

\bibitem{IM5} H.Itoyama and A.Morozov, hep-th/0211259.

\bibitem{Chetal} L.Chekhov, A.Marshakov, A.Mironov, D.Vassiliev,
hep-th/0301071.

\bibitem{WDVV}  E.Witten, 
{\it Nucl.Phys.} {\bf B340} (1990) 281; 
{\it Surveys Diff.Geom.} {\bf 1} (1991) 243; \\
R.Dijkgraaf, E.Verlinde and H.Verlinde, {\it Nucl.Phys.} {\bf B352} (1991) 59;\\
B.Dubrovin, {\it Lecture Notes in Math.} {\bf 1620},
Springer, Berlin, 1996, 120-348, hep-th/9407018; \\
E.Getzler, alg-geom/9612004.

\bibitem{MMM} A.Marshakov, A.Mironov and A.Morozov,
{\it Phys.Lett.} {\bf B389} (1996) 43-52, hep-th/9607109; \\
{\it Mod.Phys.Lett.} {\bf A12} (1997) 773-788, hep-th/9701014;\\
{\it Int.J.Mod.Phys.} {\bf A15} (2000) 1157-1206, hep-th/9701123.

\bibitem{IM6} H.Itoyama and A.Morozov, hep-th/0212032.

\bibitem{prepth} 
H.Itoyama and A.Morozov, {\it Nucl.Phys.} {\bf B491} (1997) 529, hep-th/9512161;\\
A.Gorsky, A.Marshakov, A.Mironov and A.Morozov, 
{\it Nucl.Phys.} {\bf B527} (1998) 690-716, hep-th/9802007;\\
J.Edelstein, M.Marino and J.Mas, {\it Nucl.Phys.} {\bf B541} (1999)
671, hep-th/9805172; \\
J.Edelstein and J.Mas, hep-th/9902161.

\bibitem{NeF}
N.Nekrasov, hep-th/0206161; \\
R.Flume and R.Poghossian, hep-th/0208176.

\bibitem{loop}
F.David, {\it Mod.Phys.Lett.} {\bf A5} (1990) 1019;\\
J.Ambjorn, J.Jurkiewicz and Yu.Makeenko, {\it Phys.Lett.} {\bf 251B} (1990) 517;\\ 
 A.Mironov and A.Morozov, {\it Phys.Lett.} {\bf 252B} (1990) 47;\\
H.Itoyama and Y.Matsuo, {\it Phys.Lett.} {\bf 255B} (1991) 202;\\
F.Cachazo, M.Douglas, N.Seiberg and E.Witten, hep-th/0211170.

\bibitem{Mat} M.Matone, hep-th/0212253.
\bibitem{DP} A. Dymarsky, V. Pestun, hep-th/0301135.

\end{thebibliography}
\end{document}